%% file: Formatting-Instructions-LaTeX-2021.tex
\def\year{2021}\relax
\documentclass[journal]{IEEEtran}
\usepackage{times}  
\usepackage{helvet} 
\usepackage{courier}  
\usepackage[hyphens]{url}  
\usepackage{graphicx} 
\usepackage{caption} 
\usepackage{amssymb}
\usepackage{enumerate}
\usepackage{subfigure}
\usepackage{amsmath}
\usepackage{amsthm}
\usepackage[linesnumbered,ruled, noend]{algorithm2e}

\usepackage{multirow}
\usepackage{cases}
\usepackage{textcomp}
\usepackage{booktabs}

\usepackage[switch]{lineno}

\title{Who Are the `Silent Spreaders’?:\\Contact Tracing in Spatio-Temporal Memory Models}
\author{
	Yue~Hu,
	Budhitama~Subagdja,
	Ah-Hwee~Tan,~\IEEEmembership{Senior~Member,~IEEE,}
	Chai~Quek,~\IEEEmembership{Senior~Member,~IEEE,}
	and~Quanjun~Yin
	\thanks{Y. Hu and Q. Yin are with the College of Systems Engineering, National University of Defense Technology, Changsha, Hunan, 410073 China (e-mails: huyue.cse@gmail.com; yin\_quanjun@163.com). This work was done during a visit by the first author to Nanyang Technological University, Singapore.}
	\thanks{B. Subagdja and A.-H. Tan are with the School of Information Systems, Singapore Management University, while on leave from the School of Computer Science and Engineering, Nanyang Technological University, Singapore (emails: budhitamas@smu.edu.sg and ahtan@smu.edu.sg).}
	\thanks{C. Quek is with the School of Computer Science and Engineering, Nanyang Technological University, 639798 Singapore (e-mail:ashcquek@ntu.edu.sg.}
}
\begin{document}

\maketitle

\begin{abstract}
The COVID-19 epidemic has swept the world for over a year. 
However, a large number of infectious asymptomatic COVID-19 cases (\textit{ACC}s) are still making the breaking up of the transmission chains very difficult.
Efforts by epidemiological researchers in many countries have thrown light on the clinical features of ACCs, but there is still a lack of practical approaches to detect ACCs so as to help contain the pandemic.
To address the issue of ACCs, this paper presents a neural network model called Spatio-Temporal Episodic Memory for COVID-19 (\textit{STEM-COVID}) to identify ACCs from contact tracing data.
Based on the fusion Adaptive Resonance Theory (\textit{ART}), the  model encodes a collective spatio-temporal episodic memory of individuals and incorporates an
effective mechanism of parallel searches for ACCs.
Specifically, the episodic traces of the identified positive cases are used to map out the episodic traces of suspected ACCs using a weighted evidence pooling method.
To evaluate the efficacy of STEM-COVID, a realistic agent based simulation model for COVID-19 spreading is implemented based on the recent epidemiological findings on ACCs.
The experiments based on rigorous simulation scenarios, manifesting the current situation of COVID-19 spread, show that the STEM-COVID model with weighted evidence pooling  has a higher level of accuracy and efficiency for identifying ACCs when compared with several baselines.
Moreover, the model displays strong robustness against noisy data and different ACC proportions, which partially reflects the effect of breakthrough infections after vaccination on the virus transmission.
\end{abstract}

\section{Introduction}

\noindent The outbreak of the novel coronavirus, severe acute
respiratory syndrome coronavirus 2 (SARS-CoV-2), has caused more than twenty million confirmed cases in the world.
In stark contrast to SARS and MERS, two previously coronaviruses caused respiratory diseases, a much larger proportion of COVID-19 patients never develop symptoms like fever and cough but can be contagious while shedding the coronavirus~\cite{wang2020characterization, he2020temporal, furukawa2020evidence}.
While presentation of symptoms facilitates case detection, the asymptomatic COVID-19 cases, or ACCs, are usually unaware of having been infected and proceed with their normal life while carrying the virus.
Therefore, many researchers call such cases as `silent spreaders’ of SARS-CoV-2~\cite{long2020clinical}.
Many countries with their current control measures only isolate patients since the onset of symptoms and potential post-symptomatic transmissions can be interrupted~\cite{he2020temporal}.
However, such efforts are ineffective if ACCs still have continual contacts with others, thereby causing transmissions.


To understand the characteristics of asymptomatic transmissions, many researchers have recently focused on issues such as the proportion of ACCs over all infected cases, the duration of virus shedding and the temporal dynamics of infectiousness~\cite{lee2020clinical,long2020clinical,hu2020clinical,bai2020presumed,qiu2020clinical,mizumoto2020estimating,tian2020characteristics}.
These works also suggest a prolongation of public health interventions.
Nonetheless, given that quick population screening for all positive cases is technically hard and financially expensive to implement, alternative proactive and cost-effective approaches are in an urgent need to recognize those silent spreaders out of the whole population.

To address the above issues, this paper proposes a collective episodic memory based computational model to identify ACCs by learning and reasoning over the spatio-temporal trajectories of a population.
The basic consideration is that people sharing the same space at the same time have a chance of infecting each other. 
Therefore, if one has appeared at the same places at the same periods as many others who are later tested positive with the coronavirus, he/she could be considered as a possible source of transmissions.

Episodic memory, as a form of long-term memory, is a record of sequential events associated with contextual information, e.g., observations, activities, emotions, times and places.
It enables a cognitive system to develop adaptive behaviors by recalling past experiences.
For an efficient encoding and recalling mechanism, this work extends a Spatio-Temporal Episodic Memory (STEM) model~\cite{chang2017encoding} to encode the spatio-temporal traces and the tested positivity of SARS-CoV-2 of different individuals.
The building block of the model is a class of biologically-inspired self-organized neural networks, known as fusion Adaptive Resonance Theory (ART) networks~\cite{tan2007intelligence, tan2019self}. 
With the inherited ART properties~\cite{carpenter1987massively}, STEM-COVID is designed to learn different types of cognitive nodes, generalize across events experienced by different individuals and aggregate the events of an individual into a higher-level cognitive code.

Based on the encoded collective episodic traces of a population, the STEM-COVID model can be applied in a two-step algorithm for ACC search.
The first step is to pool the episodic traces of all tested positive and symptomatic COVID-19 cases ($t$-SCCs) and use the pooled memory trace to search among the untested cases for individuals having similar episodic memory trace. 
Those cases with higher similarities are thus considered to have a higher likelihood to be asymptomatic spreaders of the SARS-CoV-2 virus.
Owing to the incremental and real-time learning dynamics of the fusion ART networks, both the encoding and retrieval operations can be efficiently performed.

It is important to note that contact tracing for ACCs is a new problem, different from other contact tracing applications. 
While there have been seemingly similar systems proposed for contact tracing, they serve to provide warnings to individuals who share the same time and space with positively identified COVID-19 cases so they can be wary of the possibility of being infected~\cite{cho2020contact}. 
In contrast, the task studied in this paper is to identify hidden asymptomatic cases who may have infected those positively identified cases in the first place.

In summary, the main novelties and technical contributions of this paper include: (a) a spatio-temporal data simulation model of COVID-19 spreading designed based on recent epidemiological findings; (b) an extended spatio-temporal episodic memory model based on self-organizing neural networks for encoding and associating the collective episodic traces of individuals with association to their respective current status of COVID-19 positivity; (c) an efficient search process for ACCs by pooling the episodic traces of positive COVID-19 cases and searching over untested cases in parallel.  

For performance evaluation, empirical experiments are conducted by running the simulation model under multiple scenarios with different degrees of infection and population sizes. 
Our experimental results show that STEM-COVID is able to identify ACCs with a reasonably high level of accuracy and efficiency.


\section{Problem Formulation}
\label{sec_problem}

This paper aims to provide a solution to ACC identification from a perspective of cognitive computation.
The approach taken is to model the spatio-temporal trajectories of different individuals in a collective manner.

Consider a population size of $N$ individuals, each spatio-temporal data point of an individual can be represented by an \emph{event} in the form of $\varepsilon = (t, p)$,  where $t$ is a time stamp in the unit of hours and $p$ is the ID of one place.
We only consider a limited spatial and temporal scale, i.e. $0\leq t<T$ and $0\leq p<P$ where $T$ is the the whole duration during which our data is collected and $P$ is the number of all places.
The \emph{episodic trace} of an individual is then given by a sequence of events as $e=<\varepsilon_0, \varepsilon_1, ..., \varepsilon_{T'}>$.
Among all the individuals, some are symptomatic COVID-19 patients who are diagnosed positive with the virus and isolated after the symptom onset, i.e. $t$-SCCs.
The lengths of traces of such individuals could be shorter than $T$ due to potential halfway isolation.
Others remain untested given no occurrence of symptoms, among whom some are ACCs, some are SCCs at their pre-symptomatic stage and the rest are healthy people.
Therefore, the COVID-19 positivity of an individual with index $i$ can be $CP_i \in \{0, 1\}$ where $1$ indicates a positive and symptomatic patient while an untested individual is labeled as $CP_i = 0$.

Given the episodic traces of all individuals and their respective $CP_i$ labels, the task is to identify that how possible each untested individual is an ACC.

\section{Proposed Approach: Spatio-Temporal Data Modeling and Reasoning}
\label{sec_approach}

In this section, we first review the basics of fusion ART and the STEM model.
Next, the architecture of the STEM-COVID model is presented, followed by the algorithms for collective episodic memory encoding and ACC search.
Finally, an analysis of computation complexity is provided.

\subsection{Fusion ART Networks}
\label{sec_sec_fusion_art}

\begin{figure}[!t]
	\centering
	\includegraphics[width = 0.9\linewidth]{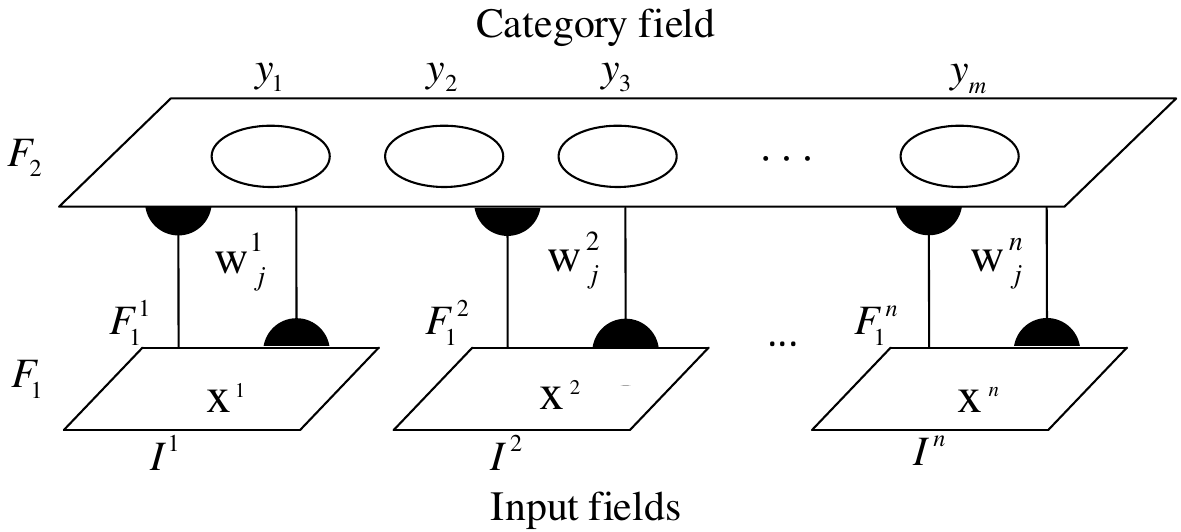}
	\caption{
		The architecture of fusion ART.
	}
	\label{fig:fusion_ART}
\end{figure}

Adaptive Resonance Theory (ART) is a neural network architecture that conducts bi-directional bottom-up and top-down pattern matching to continuously categorize the input patterns~\cite{carpenter1987massively}.
Fusion ART is a form of multi-channel ART~\cite{tan2007intelligence, tan2019self}, as shown in Fig.~\ref{fig:fusion_ART}.

%
%
%

Let $\mathbf{I}^k = (I^k_1, I^k_2, ..., I^k_{nk})$ and $\mathbf{I} = (\mathbf{I}^1, \mathbf{I}^2,...,\mathbf{I}^n)$ be an input vector and the concatenation of input vectors of all channels, respectively, where $I^k_i \in [0,1]$.
Let $F^k_1$, $\mathbf{x}^k$, $\mathbf{y}$, and $\mathbf{w}^k_j$ be the $k$th input field in $F_1$, the activity vector after receiving $\mathbf{I}^k$, the activity vector of $F_2$, and the weight vector of the $j$th node in $F_2$ for learning the input in $F^k_1$, respectively.

\textit{Parameters}: Each field's dynamics is determined by its choice parameter $\alpha^k \geq 0$, learning rate $\beta^k$, contribution parameter $\gamma^k$ and vigilance $\rho^k$, where $\beta^k, \gamma^k, \rho^k \in [0, 1]$.

The operations for fusion ART dynamics are briefly described as follows.
A more detailed description can be found in~\cite{tan2019self}.

The \textit{Code Activation} process on the $j$th node in $F_2$ is controlled by the choice function:
\begin{equation}\label{eq:choice_func}
T_j = \sum_{k=1}^{n}\gamma^k \frac{|\mathbf{x}^k\wedge \mathbf{w}^k_j|}{\alpha^k+|\mathbf{w}^k_j|}.
\end{equation}

A \textit{Code Competition} process selects an $F_2$ node $J$ with the highest $T_j$ followed by a \textit{Template Matching} to check if resonance occurs so that:
\begin{equation}\label{eq:template_matching}
m^k_J=\frac{|\mathbf{x}^k \wedge \mathbf{w}^k_J|}{|\mathbf{x}^k|} \geq \rho^k, 1\leq k \leq n.
\end{equation}

If no selected node in $F_2$ meets the vigilance, an uncommitted node is recruited in $F_2$ as a new category node.

A \textit{Template Learning} process is applied to the connection weights once resonance occurs. 
\begin{equation}\label{eq:template_learning}
\mathbf{w}^{k(\text{new})}_J = (1-\beta^k)\mathbf{w}^{k(\text{old})}_J + \beta^k(\mathbf{x}^k \wedge \mathbf{w}^{k(\text{old})}_J).
\end{equation}

The selected node $J$ can be readout to the the corresponding $F^k_1$ field by $\mathbf{x}^{k(\text{new})} = \mathbf{w}^k_J$ as the output.

\subsection{The STEM Architecture}
\label{sec_sec_STEM}


\begin{figure}[!t]
	\centering
	\subfigure[STEM]{\label{fig:STEM}\includegraphics[width = 0.8\linewidth]{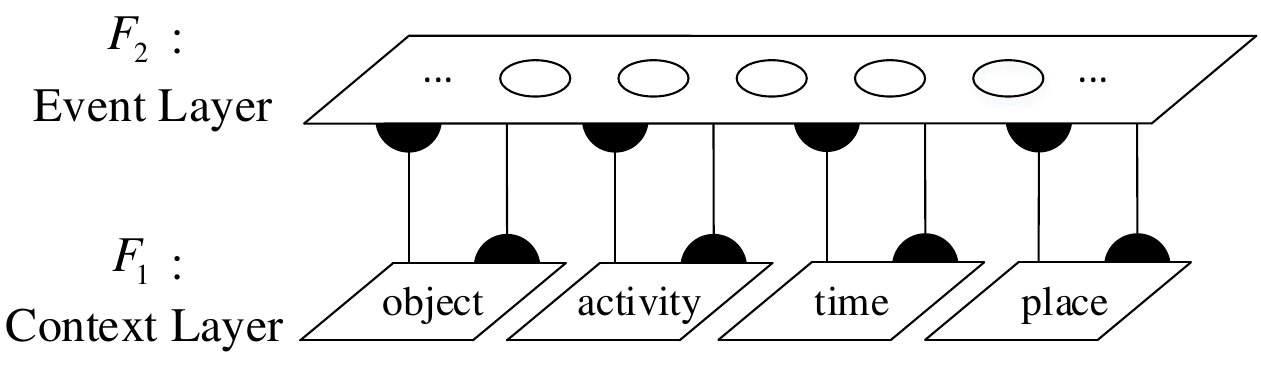}}
	\subfigure[STEM-COVID]{\label{fig:STEM-COVID}\includegraphics[width = 0.9\linewidth]{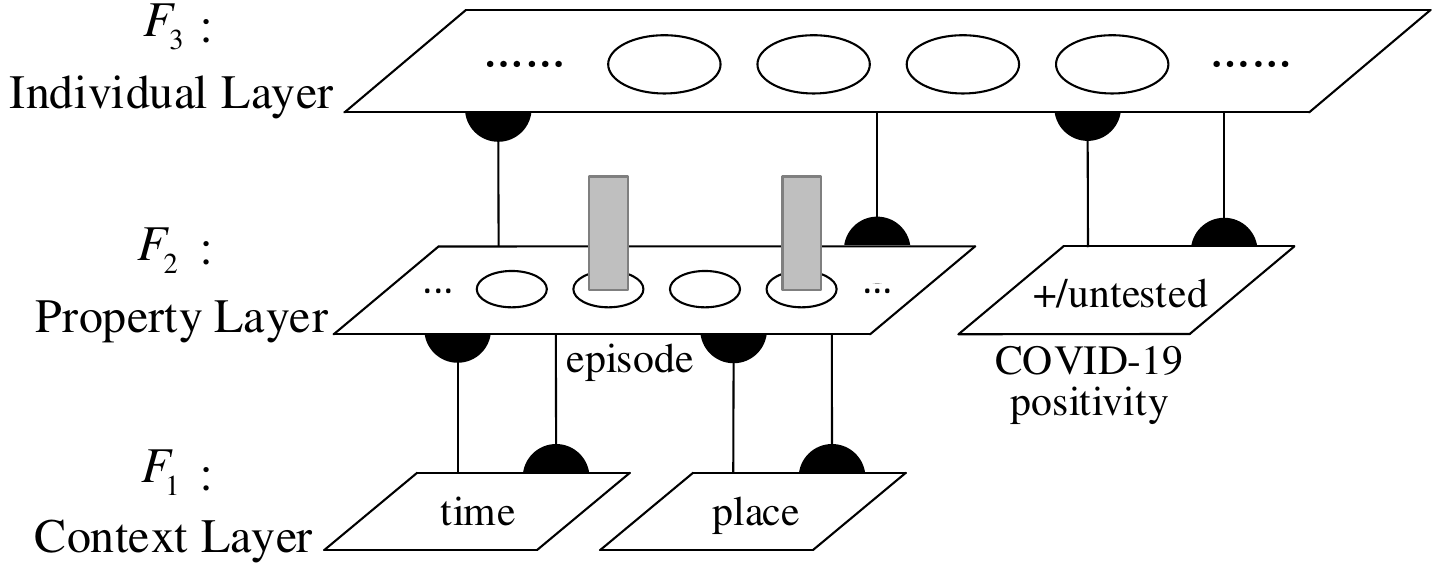}}
	\caption{
		The architectures of STEM and STEM-COVID.
	}
	\label{fig:STEMs}
\end{figure}

Previous models of episodic memory include Episodic Memory-ART (EM-ART)~\cite{wang2012neural, subagdja2015neural} and STEM~\cite{chang2017encoding}.
While EM-ART learns a hierarchical encoding scheme of sequential events and episodes, STEM explicitly represents the time and space of individual events without an episode layer.

As shown in Fig.~\ref{fig:STEM}, the original STEM architecture includes four input fields to represent the context of an event including the \emph{time} and \emph{place}.
The place property of an event in STEM is characterized by the real-world coordinate and landmark specification.
Each category node in the $F_2$ layer represents an event in response to the contextual information presented in the $F_1$ layer.

\subsection{The STEM-COVID Architecture}
This paper combines the structures of EM-ART and STEM by adding an \emph{individual} layer on top of STEM to collect the properties of different individuals.
Fig.~\ref{fig:STEM-COVID} shows the overall architecture of this new STEM-COVID network.
The structure can be viewed as two two-channel fusion ART networks stacked in a hierarchical manner, similar to EM-ART~\cite{wang2012neural, subagdja2015neural}.
In the application of contact tracing, only the time and place fields are needed in the context layer.

As an intermediate representation, the \emph{episode} field serves as both a category field of the bottom fusion ART network and an input filed of the top one.
After the events experienced by an individual are all learned from the bottom network, our encoding mechanism will form an episode vector $\mathbf{E}$ with a length equal to the number of all currently existing event nodes in the episode field.
In this binary vector, values of 1s correspond to events experienced by the individual while the rest are zeros.
Besides the episode field, another field indicates the COVID-19 positivity of the individual $i$ by a vector $\mathbf{c}=(CP_i)$.
Based on $\mathbf{E}$ and $\mathbf{c}$, the top network learns a unique code for each individual in the $F_3$ layer.

\subsection{Collective Episodic Memory Encoding}
\input{alg_episodic_learning}

As mentioned, episodic memory considers an individual episode as a sequence of events as $e=<\varepsilon_0, \varepsilon_1, ..., \varepsilon_{T'}>$, where an event is formalized as $\varepsilon_l = (t_l, p_l)$.
We present a hierarchical scheme for constructing collective episodic memory based on STEM-COVID in Alg.~\ref{alg:episodic_learning}.

%
The algorithm consists of a loop through episodic traces of all individuals in the considered population (Line~\ref{alg:episodic:individual_start}-\ref{alg:episodic:individual_end}).
After an episode vector $\mathbf{E}$ is formed for each trajectory through learning in the interaction between the $F_1$ and $F_2$ layers (Line~\ref{alg:episodic:episode_start}-\ref{alg:episodic:episode_end}), the top network creates a unique cognitive code in the $F_3$ layer for each individual based on $\mathbf{E}$ and the positivity vector $\mathbf{c}$ presented in $F_2$ (Line~\ref{alg:episodic:top_start}-\ref{alg:episodic:top_end}).
In this application, it suffices to set $\mathbf{E}$ as a binary vector to indicate the events of one individual (Line~\ref{alg:episodic:top_start}), since we will not exploit the temporal ordering of events within an episode.
In contrast, EM-ART encodes the ordering in a sequence by a gradient encoding method~\cite{subagdja2015neural}.

To encode an event, the algorithm first normalizes the time and place attributes into real values $t', p' \in [0,1)$ (Line~\ref{alg:episodic:normalize}) and then present the input to the input fields in the $F_1$ layer (Line~\ref{alg:episodic:presentF1}).
A standard learning process in fusion ART networks involves resonance search wherein nodes in $F_2$ are visited through some iterations of code competition and template matching.
However, since an event can be considered unique with a distinct time step and place ID, the resonance search can be simplified by employing template matching only once (Line~\ref{alg:episodic:resonance_event_start}-\ref{alg:episodic:resonance_event_end}).
A new uncommitted event code will be recruited for encoding the novel input, once the template matching fails (Line~\ref{alg:episodic:new_event_node}).

As the entire trajectory and positivity status are formed as $\mathbf{E}$ and $\mathbf{c}$ in $F_2$, the information about the individual can be encoded and stored as a node in the $F_3$ layer (Line~\ref{alg:episodic:top_start}-\ref{alg:episodic:top_end}).

\subsection{ACC Identification based on STEM-COVID}

Based on the collective episodic memory in the STEM-COVID model, a two-step process is proposed to identify ACCs based on the episodic traces of the known SCCs.

\input{alg_acp_detection}

\subsubsection{Evidence pooling}
The ACC identification process begins with pooling the episodic traces of all $t$-SCCs into an evidence vector $\mathbf{E}$, as shown in Alg.~\ref{alg:acp_detection} (Line~\ref{alg:acp:merge_start}-\ref{alg:acp:merge_end}).
All nodes representing $t$-SCCs can be activated by choice function wherein the positivity vector $\mathbf{c}$ is set to (1) while setting the contribution parameters $\gamma^c=1$ and $\gamma^e=0$ ignoring the episodic trace (Line~\ref{alg:acp:gamma_merge}-\ref{alg:acp:activate_positive}). 
For codes with positive activation values in $F_3$, the values are readout to the episode field in $F_2$ and integrated using a fuzzy \emph{OR} operation (Line~\ref{alg:acp:unifying_vector_start}-\ref{alg:acp:merge_end}). 

\subsubsection{ACC search}
Searching for ACCs can be done by computing the similarities between this combined trace and the traces of untested individuals. 
Accordingly, the contribution parameters are set to $\gamma^e=\gamma^c=1$ and the positivity vector $\mathbf{c}$ is set (0) to search for untested individuals (Line~\ref{alg:acp:similarities_start}-\ref{alg:acp:set_untested}).
With the input vectors $\mathbf{E}$ and $\mathbf{c}$, every single node in the $F_3$ layer will be activated by the choice function which is in essence an evaluation of the similarity between the input pattern and the cognitive code (Line~\ref{alg:acp:activate_unstested_start}-\ref{alg:acp:activate_unstested_end}).

A higher activation value obtained in this step indicates that the corresponding untested individual has experienced more events overlapped with those of all $t$-SCCs.
The potential ACCs can be identified by selecting the nodes in $F_3$ with activation values greater than a threshold $\delta_c$ (Line~\ref{alg:acp:similarities_end}).
Another strategy is to select top $k$ individual nodes based on the activation values where $k$ is a user-defined parameter.

\subsection{Complexity Analysis}
\label{sec_sec_complexity}

This subsection presents an analysis of the space and time complexity of the episodic memory construction and ACC search algorithm.
For space complexity, we are only concerned about the number of event nodes learned in the episode field, given that the number of nodes in the $F_3$ layer is equivalent to a constant $N$.
The time complexity of learning an event is also worth an analysis.
Moreover, we compare the time complexity of ACC detection between STEM-COVID and a baseline.

\subsubsection{Space complexity of STEM-COVID}

The worst-case space complexity occurs when all individuals share no common events so that the number of event nodes can be $T\cdot N$.
However, there can only be $T\cdot P$ combinations of time and place values.
Therefore, the worst-case space complexity for episode field is $T\cdot \text{min}(N, P)$.

\subsubsection{Time complexity of STEM-COVID}

In the standard fusion ART model, the worst-case time complexity for learning a category node is $O(mn^2)$ if no matched node is found, where $m$ is the number of attributes in the input fields and $n$ is the number of existing category nodes~\cite{chang2017encoding}.
The quadratic component is incurred by the repeated code competition and template matching for resonance search.
However, we substitutes a loop of top-down match through all existing event nodes for such iterations (Line~\ref{alg:episodic:resonance_event_start}-\ref{alg:episodic:resonance_event_end} in Alg.~\ref{alg:episodic_learning}).
In virtue of this simplification, the worst-case time complexity is reduced to $O(mn_{\varepsilon})$ where $n_{\varepsilon}$ is the number of existing event nodes and $m=2$ given the two single-attribute fields in $F_1$.
With no matching required, the time complexity of learning in the top network is just $O(n_{\varepsilon})$ where $n_{\varepsilon}$ serves as the number of input attributes.

As for the time complexity for ACC identification, while the evidence pooling involves $N$ times of calling the choice function (Line~\ref{alg:acp:activate_positive} in Alg.~\ref{alg:acp_detection}) and at most $N$ times of activity readout and fuzzy \emph{OR} operations (Line~\ref{alg:acp:unifying_vector_start}-\ref{alg:acp:merge_end}), the similarity computation only requires activating $N$ nodes in the $F_3$ layer (Line~\ref{alg:acp:activate_unstested_end}).
Summed over all these steps, the time complexity is $O(N)$ or linear.

\input{alg_baseline}

\subsubsection{Time complexity of baseline algorithm}

To show the efficiency of our algorithm for ACC detection, we present a naive algorithm as a baseline for computing the similarity between the unified events of $t$-SCCs and the trace experienced by each untested individual.
As shown in Alg.~\ref{alg:baseline}, the baseline simply counts the number of events commonly experienced by one untested individual and any $t$-SCC and considers the ratio of this count to the number of all events experienced by the untested individual as a metric of similarity (Line~\ref{alg:baseline:count_start}-\ref{alg:baseline:similarity}).
The whole process applies three nested loops, thus involving $N_{u} \cdot T \cdot n_{t}$ times of event matching where $n_{t}=|\{\varepsilon_{t}\}|$ and at most equals to $N_t \cdot T$.
Hence the worst-case time complexity of this baseline is $O(N^2T^2)$ where $N=N_{t}+N_{u}$, which is quadratic compared to STEM-COVID.

\section{COVID-19 Data Simulation}
\label{sec_simulation}

For empirical performance evaluation, a simplified simulation model of the real world, where no public health interventions are exerted, is designed to emulate the spreading of coronavirus in communities.
This section will briefly introduce the spatial representation, the life cycle of individual agents, the temporal dynamics of virus transmissibility and the key parameter settings in order.

\subsection{Spatial Representation}

\begin{figure}[!t]
	\centering
	\includegraphics[width = 0.8\linewidth]{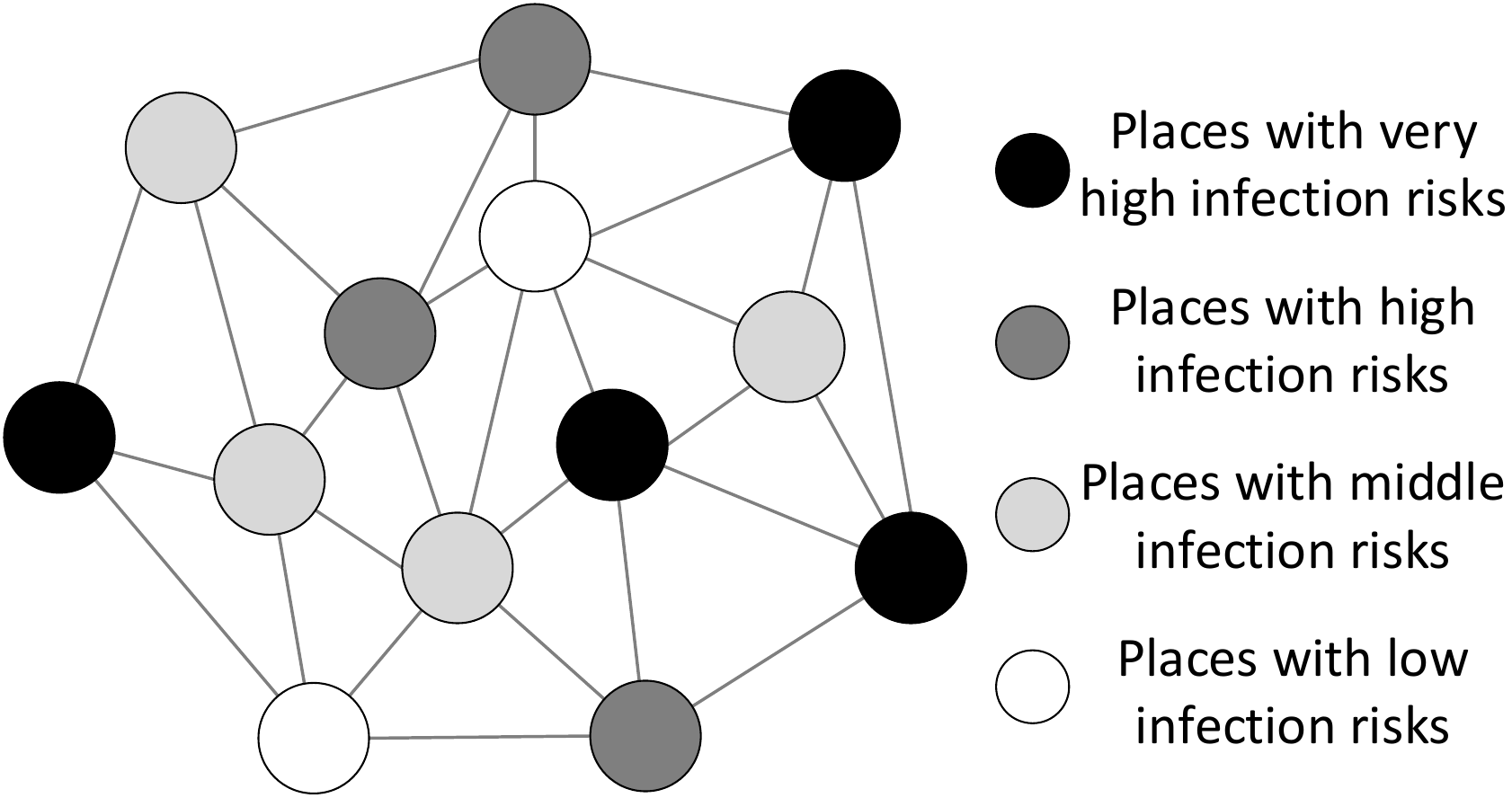}
	\caption{
		A graphic representation of the spatial model.
	}
	\label{fig:graph}
\end{figure}

We discretize the simulated environment into a network of places that are reachable from each other.
As shown in Fig.~\ref{fig:graph}, the neighboring places can be connected by walkways or roads, and the whole space can be seen as a graph.

People undergo different levels of risks to be infected by a coronavirus carrier at different kinds of places, typically influenced by the crowd density, the degree of ventilation and other factors.
This is the reason why lots of transmission chains are found in family clusters~\cite{bai2020presumed, tong2020potential}.
In view of such findings, we categorize the places into different groups.
Places with very high infection risks are mainly homes, nursing rooms, etc.
Places such as cafes, restaurants and shopping malls are highly risky for consumers to be infected.
Workplaces like offices and factories can be considered with middle risks and open spaces like playgrounds and parks carry low risks to people.

\subsection{Life Cycle of Agents}


To simulate the dynamic activities of individuals between different places, we model each individual as an agent who is assigned a daily life cycle,
whereby they stay at various kinds of places for different times.
For example, the agents stay at homes (very high-risk places) for about 10 hours, work at offices (middle-risk) for 8 hours, go shopping or socialize (high-risk) for 3 hours and do outdoor activities (low-risk) for 3 hours.
The durations vary from day to day and from person to person.
Though more possibilities of life styles exist in reality, our assumption makes sense because it represents a typical mode for many people.

\begin{figure}[!t]
	\centering
	\includegraphics[width = 0.9\linewidth]{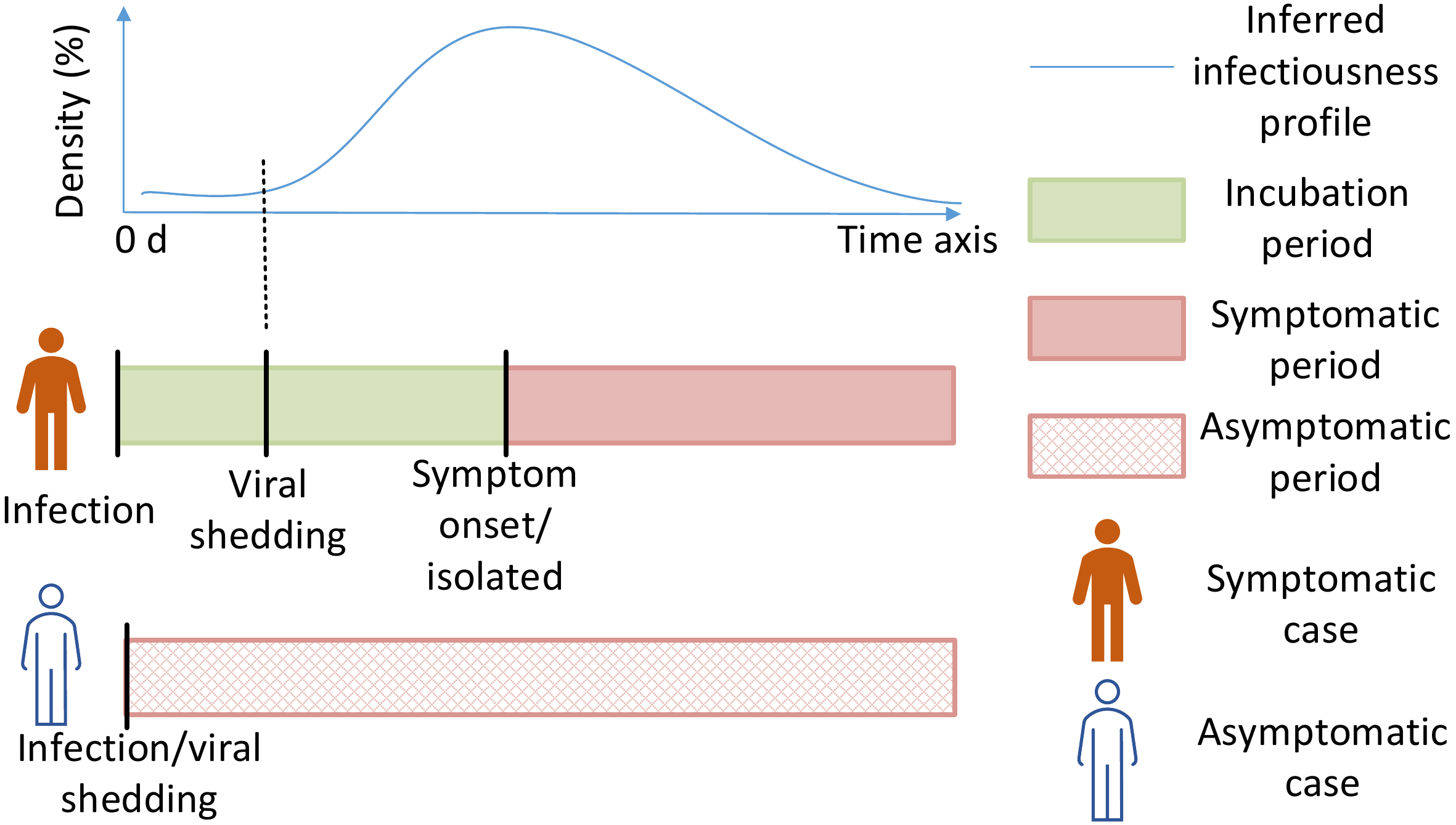}
	\caption{
		The temporal dynamics of SCCs and ACCs.
	}
	\label{fig:temporal_dynamics}
\end{figure}

Along with the transformations between locations, a simulation of virus spreading requires some additional consideration for the temporal dynamics in clinic features of the infected people.
As many studies reported~\cite{bai2020presumed, he2020temporal, long2020clinical}, SCCs are infectious not only after symptom onset, but also at their pre-symptomatic stage.
We illustrate the dynamics in Fig.~\ref{fig:temporal_dynamics}, which is adapted from~\cite{he2020temporal}.
SCCs have low infectiousness at the start of their incubation period (time from infection to symptom onset), which climbs quickly to a highest level around symptom onset.
Based on the clinic researches, this paper simplifies the infectiousness profile of SCCs by a step function, whereby they are not contagious from infection until substantial viral shedding begins, and then keep consistent infectiousness until the onset of symptoms and isolation.

In contrast, some investigations found significantly longer duration of viral shedding in ACCs than SCCs, e.g., an interquartile range of 15-26 days, even longer than the total time in our simulation~\cite{long2020clinical}.
Therefore, we assume that ACCs are constantly contagious throughout the simulation once they are infected with the virus.

\subsection{Parameter Settings}


\subsubsection{Incubation period}
A number of researches have reported different estimations of the duration of incubation periods of SCCs~\cite{li2020early, backer2020incubation, lauer2020incubation, linton2020incubation}.
Most of them agree that the median duration is around 5 days and the durations of most cases fall into the range of 3-7 days.
In view of the estimations, we use a Gaussian distribution $T_{in}=(T_1+T_2)\sim \mathcal{N}(120h, (12h)^2)$ to sample a duration $T_{in}$ of incubation period for each SCC.
As mentioned, the duration consists of two parts, i.e. $T_1=0.2\cdot T_{in}$ for time from infection to viral shedding when the patient is not infectious, and $T_2$ from viral shedding to symptom onset.

\subsubsection{Proportion of ACCs}
The existing estimations of the proportion of asymptomatic cases over all infections can be varied from 5\% to 80\%~\cite{carl2020covid, WHO2020differences}. 
But a large number of them fall into the range from 10\% to 30\%, such as studies on cases in Zhejiang, China, the Diamond Princess cruise ship and Singapore~\cite{qiu2020clinical, mizumoto2020estimating,chew2020almost}.
In view of these epidemiological findings, this paper sets a percentage 20\% of all infections as ACCs.

\subsubsection{Infection rates}
We set infection rates in a unit of hours for the mentioned four kinds of places as $r_{vh}=0.01, r_{h}=0.005, r_{m}=0.001$ and $r_{l}=0.0001$ respectively.
Note that these rates keep the same no matter the source of transmissions is symptomatic or asymptomatic, in view of the similar viral load found in the different kinds of cases~\cite{lee2020clinical}.
As such, if an individual in a household is infectious, other families would be infected during the 10 hours of stay with a probability of $p(\text{infected}|vh)=1-(1-r_{vh})^{10}\approx 9.6\%$.
Similarly, the infection probabilities at other places are $p(\text{infected}|h)=1-(1-r_{h})^3=1.5\%$, $p(\text{infected}|m)=1-(1-r_{m})^8=0.8\%$ and $p(\text{infected}|l)=1-(1-r_{l})^3=0.03\%$.
With $k>1$ coronavirus carriers at the same space, we simply increase the infection probabilities to $\text{min}\{1.0, p(\text{infected})\cdot k\}$.
We will verify these settings through comparing the results of simulation and recent epidemiological findings.

\section{Experiments}
\label{sec_experiments}

This section will report the performance of our approach in terms of the effectiveness and efficiency for identifying ACCs.
The settings of three different scenarios and performance measures will be first described, before the quantitative results and discussion are presented. 

\subsection{Simulation Scenarios}

\input{tab_scenarios}

All simulations run for twenty days, i.e. $T=480$ hours.
For simplicity, every four agents are clustered in one household while extension to more complex family structures is easily applied in the model.
The simulation model is run under three different scenarios, whose key parameters are listed in Table.~\ref{tab:scen}.
Besides the population sizes $N$  and the numbers of different kinds of places, $N_{u\_0}$ agents are randomly selected as the index cases who are asymptomatic and infectious from the very beginning of the simulation, also called as `patient zero'.
The scenario S200H differs from S200N by an index case whose life cycle is modified, so that the agent will everyday stay about ten hours at high-risk places, marked by `H'.
We design such a scenario to simulate superspreading events where a patient infects many others in crowed indoor gatherings~\cite{aylin2020coronavirus}.
The scenario S1000N, with a population size and spatial scale fivefold as those in S200N, is used to investigate the effect of population size on performance.
The simulation model runs 15 times under each scenario with different random seeds to generate multiple simulation instances for statistical results.
%


\subsection{Performance Measures}
A detection is considered successful if one of the $k$ agents identified with the largest trace similarities is a true ACC.
The ratio of the number of successful simulation instances to the total, as a \emph{top-$k$ success rate}, is then used to demonstrate the detection accuracy.
Beyond normal ACCs, it will help more if the index cases can be identified, since they could be superspreaders who infected many others.

Recall that we have theoretically compared the time complexity of STEM-COVID and the baseline for ACC search.
More practically, we will present the time costs required by them to complete the computation of trace similarities in the simulations.
All the experiments are run on a laptop with Intel(R) Core(TM) i7-9750H CPU @ 2.60GHz.

\subsection{Results and Discussion}
Before looking into the performance on ACC detection, we first summarize the results of simulations. 
Finally, we will compare the time costs of STEM-COVID and the baseline.
\subsubsection{Summary of simulations}

\begin{figure}[!t]
	\centering
	\includegraphics[width = 0.9\linewidth]{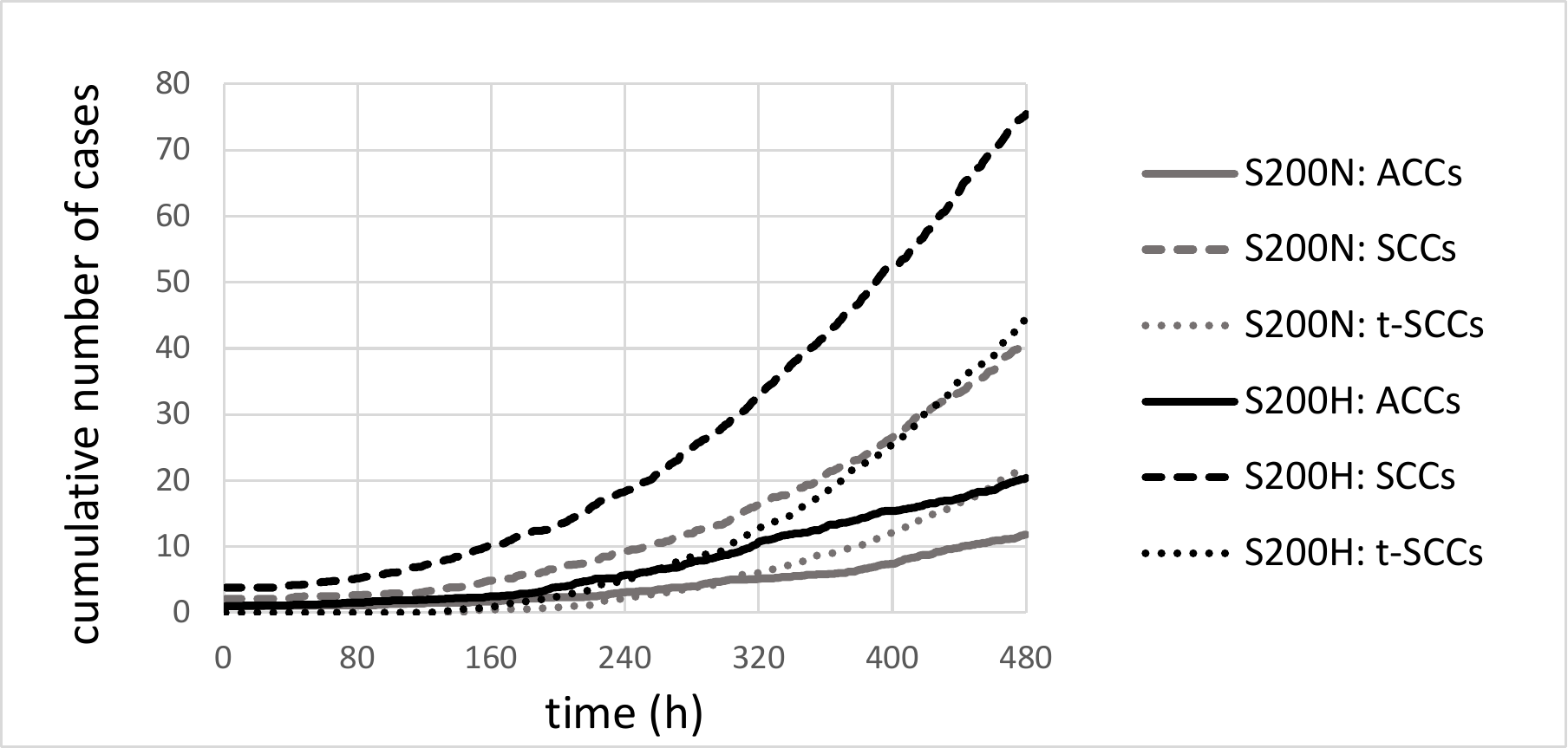}
	\caption{
		Growth of COVID-19 cases in S200N and S200H scenarios.
	}
	\label{fig:cases_overtime}
\end{figure}

Fig.~\ref{fig:cases_overtime} shows the growth of ACCs, $t$-SCCs and all SCCs over the simulation period, averaged over all 15 simulations.
The exponential curves imply a rapid increase of cases.
Notice that the S200N and S200H scenarios take about 160 hours to increase the numbers of SCCs from 20 to 40 and from 35 to 70 respectively.
The numbers tally very well with many epidemiological investigations on early transmissions of COVID-19 that its basic reproductive number ($R_0$) is over 2 and the doubling time of the cases is around 7 days~\cite{li2020early, wu2020nowcasting, volz2020report}.

Moreover, even though we change the life cycle of only one agent in S200H, the striking contrast of case numbers between S200H and S200N manifests that indoor gatherings of large numbers of people can easily drive the epidemic out of control.
Besides, the average cumulative numbers of ACCs, $t$-SCCs and SCCs in S1000N, reach 63, 121 and 225 respectively, roughly five times of those in S200N.

\begin{figure}[!t]
	\centering
	\includegraphics[width = 0.95\linewidth]{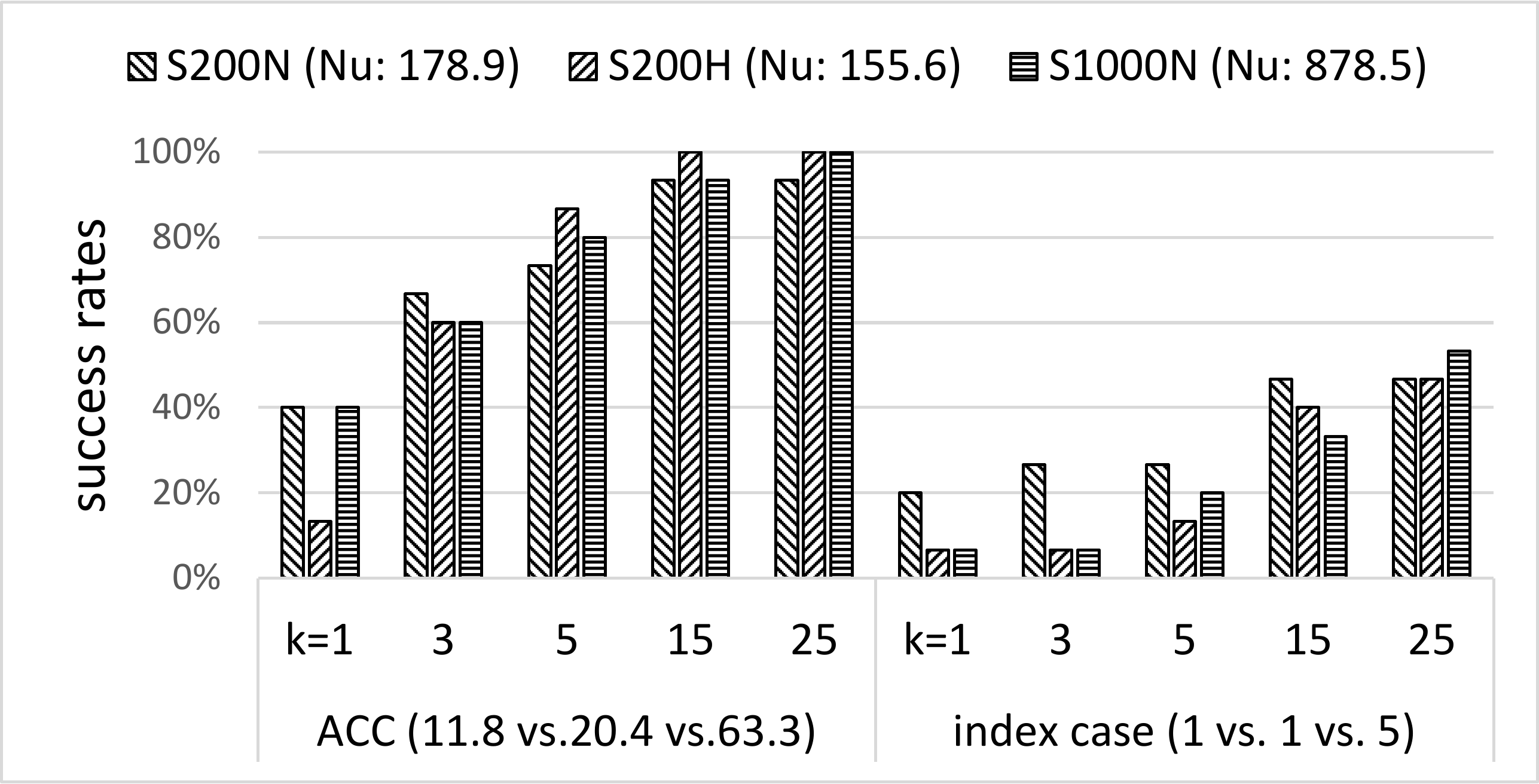}
	\caption{
		Top-$k$ success rates of the STEM-COVID model.
	}
	\label{fig:success_rate_scen1_2_3}
\end{figure}

\subsubsection{Effectiveness for ACC identification}
Fig.~\ref{fig:success_rate_scen1_2_3} shows the top-$k$ identification success rates of STEM-COVID in different scenarios, where the mean numbers of untested cases, ACCs and index cases averaged over all simulation instances are reported in corresponding brackets.
Overall, our approach reaches a reasonably high level of success rates in all scenarios to detect ACCs even with the top-3 measure, given that there are only 11.8 ACCs out of all 178.9 untested cases on average in S200N.
Even with a smaller proportion of ACCs, the success rates are higher in S200N than S200H when $k<5$  individuals are identified.
A plausible explanation is that fewer times of contacts are required to cause the same number of infections in S200H than required in S200N, so it seems more difficult for contact count based approach to detect the sources of infections.
This speculation can be verified by the fact that the index cases in S200H infected 12.8 persons on average, in contrast to only 5.0 in S200N in the same duration of simulations.
Furthermore, the performance of the algorithm in S1000N is very close to that in S200N, implying that our approach can scale up to larger scenarios without loss of effectiveness.


Among ACCs, index cases are extremely difficult to be screened because of their small proportion over the whole population.
Nevertheless, the top-3 and top-5 success rates for the only one index case in S200N are over 20\%, even higher with $k=15$ and $25$.
Comparatively, the unique life cycle of the patient zero in S200H incurs significantly lower success rates when $k \leq 5$.
However, its results improve significantly with larger values of $k$.
Furthermore, the algorithm scales up well with a larger population, considering that the top-5, 15 and 25 success rates for S1000N are much higher than the top-1, 3 and 5 success rates for S200N.

\input{tab_time_cost}


\subsubsection{Time efficiency on ACC identification}
Note that the baseline algorithm can be considered as a brute force approach to realizing the ACC identification algorithm in STEM-COVID. 
Therefore they have similar performance in terms of the identification accuracy. 
To illustrate their difference in efficiency, we report the time costs (in seconds) of running STEM-COVID and the baseline to compute the trace similarities between untested and $t$-SCCs in Table.~\ref{tab:time}.
The results are obtained by executing the algorithms once on each simulation instance.
No matter which scenario the algorithms are performed in, our approach requires much less time to complete the computation.
While under each scenario, the time costs of the baseline vary much in different simulations, the results of our approach keep relatively consistent.
The comparison demonstrates the superiority of our algorithm on computation efficiency.

\section{Conclusion}
\label{sec_conclusion}
This paper has presented an episodic memory based neural model called STEM-COVID, which is able to encode the collective spatio-temporal data of a population and support an efficient search algorithm to identify asymptomatic COVID-19 cases.
With the disease sweeping the world, we hope that this novel and cost-effective way can help the authorities trace and control the epidemic, in addition to traditional symptom-based strategies.
Simulation based experiments have proved the effectiveness and efficiency of this episodic memory  based approach.
Serving as a starting point, this work can hopefully inspire more efforts on containing the disease through advanced techniques of computational intelligence.

Nevertheless, there remain several limitations in our work that require further research.
First, the demographic and geographic models used are still too simple and small-scale compared with the real world.
Second, the complexity of the data simulation model can be expanded in several dimensions.
For example, we currently assume a consistent infectiousness profile throughout the whole duration of viral shedding.
Third, a fairly simple fuzzy \emph{OR} operation is used to integrate the spatio-temporal episodic traces of multiple $t$-SCCs.
For future work, we would like to expand the simulation of the world model to make it more realistic and incorporate more practical models of temporal dynamics of COVID-19 transmissions.
Moreover, to deal with the modeling of super large-scale scenarios, like epidemic prevention in a metropolis of millions of people, multi-resolution and hierarchical modeling of spatio-temporal data would be helpful.


\bibliographystyle{IEEEtran}
\bibliography{asymptomatic}

\end{document}

%% file: alg_episodic_learning.tex
\begin{algorithm}[!t]
	\caption{Incremental construction of collective spatio-temporal episodic memory.}
	\label{alg:episodic_learning}
	
	\KwIn{episodes of all $N$ individuals $\{e_i|0 \leq i < N\}$}
	\KwIn{positivity of all individuals: $\{CP_i\}$}
	\KwOut{learned episodic memory in STEM-COVID}
	
	\ForEach{$e_i = <\varepsilon_0, \varepsilon_1, ..., \varepsilon_{T'}>$}
	{\label{alg:episodic:individual_start}	
		initialize a set of activated codes $Js \leftarrow \varnothing$\\
		
		\tcp{\small{learning in bottom network}}
		\ForEach{$\varepsilon_l = (t_l,p_l)$}
		{\label{alg:episodic:episode_start}
			initialize a code flag: $J \leftarrow -1$\\
			normalize: $t'\leftarrow t_l/T, p' \leftarrow p_l/P$\\\label{alg:episodic:normalize}
			present $\mathbf{I}=((t'), (p'))$ to $F_1$\\\label{alg:episodic:presentF1}
			\ForEach{code $J'$ in episode field in $F_2$}
			{\label{alg:episodic:resonance_event_start}
				\If{a top-down matching between $J'$ and $\mathbf{I}$ succeeds}
				{
					$J \leftarrow J'$\\
					break\\
				}
				\label{alg:episodic:resonance_event_end}
			}
			\If{$J == -1$}
			{
				create a new code $J$ for this new pattern\\\label{alg:episodic:new_event_node}
				expand the field as an input to $F_3$\\
			}
			$Js \leftarrow Js \cup \{J\}$\\
			\label{alg:episodic:episode_end}
		}
		\tcp{\small{learning in top network}}
		form a binary vector $\mathbf{E}$: $\mathbf{E}_j=1$, if $j \in Js$\\\label{alg:episodic:top_start}
		form a positivity vector: $\mathbf{c} = (CP_i)$\\
		present the input vectors $(\mathbf{E}, \mathbf{c})$ in $F_2$\\
		recruit an uncomitted code $J$ in $F_3$\\
		do an overwrite learning: $\mathbf{w}_J^e=\mathbf{E}$ and $\mathbf{w}_J^c=\mathbf{c}$\\
		\label{alg:episodic:top_end}\label{alg:episodic:individual_end}
	}
\end{algorithm}

%% file: alg_acp_detection.tex
\begin{algorithm}[!t]
	\caption{Evidence pooling and ACC search.}
	\label{alg:acp_detection}
	
	\KwIn{episodic memory learned in STEM-COVID}
	\KwIn{a threshold of activation value $\delta_c$}
	\KwOut{a set of identified individuals}

	\tcp{\small{Step 1: Evidence activity pooling}}
	reset $F_3$ layer\\\label{alg:acp:merge_start}
	set contribution parameters: $\gamma^e \leftarrow 0, \gamma^c \leftarrow 1$\\\label{alg:acp:gamma_merge}
	present a positivity vector in $F_2$: $\mathbf{c} \leftarrow (1)$\\
	activate $F_3$ using choice function\\\label{alg:acp:activate_positive}
	reset episode field in $F_2$\\
	initialize evidence vector $\mathbf{E} \leftarrow \mathbf{0}$\\\label{alg:acp:unifying_vector_start}
	\ForEach{code $0< T_j \leq 1$ in $F_3$}
	{
		readout node $j$ to episode field: $\mathbf{x}^{e} \leftarrow \mathbf{w}^e_j$\\
		update the evidence vector: $\mathbf{E} \leftarrow \mathbf{x}^{e} \vee \mathbf{E}$\\
		\label{alg:acp:merge_end}
	}
	
	\tcp{\small{Step 2: ACC Search}}
	set contribution parameters: $\gamma^e \leftarrow 1, \gamma^c \leftarrow 1$\\ \label{alg:acp:similarities_start}
	set a positivity vector: $\mathbf{c} \leftarrow (0)$\\\label{alg:acp:set_untested}
	present $\mathbf{E}$ and $\mathbf{c}$ in $F_2$\\\label{alg:acp:activate_unstested_start}
	activate $F_3$ using choice function\\\label{alg:acp:activate_unstested_end}
	return IDs of identified individuals $\{j|T_j>\delta_c\}$\\\label{alg:acp:similarities_end}
\end{algorithm}

%% file: alg_baseline.tex
\begin{algorithm}
	\caption{A baseline of computing trace similarities.}
	\label{alg:baseline}
	
	\KwIn{merged events $\{\varepsilon_{t}\}$ of $N_{t}$ $t$-SCCs}
	\KwIn{episodes of $N_{u}$ untested individuals}
	\KwOut{similarity of each untested individual}
	
	\ForEach{untested individual $a_{i=0,1,..., N_{u}-1}$}
	{
		intialize a counter $c \leftarrow 0$\\
		retrieve the episode $e_i=<\varepsilon_0, \varepsilon_1,..., \varepsilon_{T-1}>$\\
		\ForEach{$\varepsilon_j$ in $e_i$}
		{\label{alg:baseline:count_start}
			\ForEach{$\varepsilon_l \in \{\varepsilon_{t}\}$}
			{
				\If{$\varepsilon_j == \varepsilon_l$}
				{
					$c \leftarrow c+1$\\
					break\\
				}
			}
			\label{alg:baseline:count_end}
		}
		calculate the trace similarity $s_i \leftarrow c/T$\\\label{alg:baseline:similarity}
	}

\end{algorithm}

%% file: tab_scenarios.tex
\begin{table}[!bt]
	\centering
	\setlength{\tabcolsep}{1mm}{
		\begin{tabular}{c||c|c|c|c|c|c}
			\toprule
			
			scenarios & $N$ & $P_{vh}$ & $P_h$ & $P_m$ & $P_l$ & $N_{u\_0}$\\
			\midrule

			S200N  & 	200 &	50 &	2 &	10 &	2 &	1(N)\\
			
			S200H  & 	200	&	50 & 	2 & 10 &	2 & 1(H)\\
			
			S1000N &   1000 &  250 &   10 & 50 &   10 & 5(N)\\
			
			\bottomrule
			
	\end{tabular} }
	\vspace{-0.5em}
	\caption{
		Summary of settings in the three scenarios.}
	\label{tab:scen}
\end{table}

%% file: tab_time_cost.tex
\begin{table}[!bt]
	\centering
	\setlength{\tabcolsep}{1mm}{
		\begin{tabular}{c||cc|cc|cc}
			\toprule
			
			\multirow{2}{*}{time (s)} & \multicolumn{2}{c|}{\textbf{S200N}} & \multicolumn{2}{c|}{\textbf{S200H}} & 
			\multicolumn{2}{c}{\textbf{S1000N}}\\ 
			
			& ours & baseline & ours & baseline  & ours & baseline \\
			\midrule

			Min 	& 3.1 	& 7.3	& 3.1	& 11.3	& 76.8	& 100.0\\
			
			Mean	& 3.3	& 18.5 & 3.2	& 24.7	& 81.3	& 565.8\\
			
			Max 	& 4.4	& 29.4	& 3.4	& 33.1	& 94.0	& 784.8\\
			
			\bottomrule
			
	\end{tabular} }
	\vspace{-0.5em}
	\caption{
		Time costs for computing trace similarities.}
	\label{tab:time}
\end{table}